\documentclass[nofootinbib,floats,aps,groupedaddress,preprint,12pt]{revtex4-1}

\usepackage{amssymb,amsmath,amsbsy}
\usepackage{mathrsfs,verbatim,enumerate}
\usepackage[dvips]{graphics}
\usepackage{epsfig}
\usepackage{graphicx}
\usepackage{epsfig}
\usepackage{graphics}

\newcommand{\nn}{\nonumber}
\newcommand{\eff}{{\cal F}}
\newcommand{\emm}{{\cal M}}
\newcommand{\mb}{{\cal M}_B^2}
\newcommand{\mf}{{\cal M}_F^2}

\newcommand{\str}{\text{STr}}

\newcommand{\Muv}{M_{*}}

\newcommand{\drawsquare}[2]{\hbox{%
\rule{#2pt}{#1pt}\hskip-#2pt
\rule{#1pt}{#2pt}\hskip-#1pt
\rule[#1pt]{#1pt}{#2pt}}\rule[#1pt]{#2pt}{#2pt}\hskip-#2pt
\rule{#2pt}{#1pt}}

\newcommand{\Yfund}{\raisebox{-.5pt}{\drawsquare{6.5}{0.4}}}
\newcommand{\Ysymm}{\raisebox{-.5pt}{\drawsquare{6.5}{0.4}}\hskip-0.4pt%
        \raisebox{-.5pt}{\drawsquare{6.5}{0.4}}}
\newcommand{\Yasymm}{\raisebox{-3.5pt}{\drawsquare{6.5}{0.4}}\hskip-6.9pt%
        \raisebox{3pt}{\drawsquare{6.5}{0.4}}}
\newcommand{\Ythreea}{\raisebox{-3.5pt}{\drawsquare{6.5}{0.4}}\hskip-6.9pt%
        \raisebox{3pt}{\drawsquare{6.5}{0.4}}\hskip-6.9pt
        \raisebox{9.5pt}{\drawsquare{6.5}{0.4}}}

\newcommand{\Yadjoint}{\raisebox{-3.5pt}{\drawsquare{6.5}{0.4}}\hskip-6.9pt%
        \raisebox{3pt}{\drawsquare{6.5}{0.4}}\hskip-0.4pt
        \raisebox{3pt}{\drawsquare{6.5}{0.4}}}
\def\beq{\begin{equation}}
\def\eeq{\end{equation}}
\def\beqa{\begin{eqnarray}}
\def\eeqa{\end{eqnarray}}
\def\ifmath#1{\relax\ifmmode #1\else $#1$\fi}
\def\lsup#1{^{\lower 4pt\hbox{$\scriptstyle#1$}}}
\def\llsup#1{^{\lower 2pt\hbox{$\scriptstyle#1$}}}

\def\lsim{\mathrel{\raise.3ex\hbox{$<$\kern-.75em\lower1ex\hbox{$\sim$}}}}
\def\gsim{\mathrel{\raise.3ex\hbox{$>$\kern-.75em\low
er1ex\hbox{$\sim$}}}}

\def\eq#1{Eq.~(\ref{#1})}

\newcommand{\groebner}{{Gr\"{o}bner}}
\newcommand{\kahler}{{K\"{a}hler}}

\begin{document}


\pagestyle{empty}

\begin{center}
{\Large \bf Hunting for Dynamical Supersymmetry Breaking in Theories That S-confine }

\vspace{1cm}

{\sc Tongyan Lin$^{a,}$}\footnote{E-mail: tongyan@physics.harvard.edu}
{\sc John D. Mason$^{a,}$}\footnote{E-mail: jdmason@physics.harvard.edu}
{\sc\small and}
{\sc Aqil Sajjad$^{a,}$}\footnote{E-mail: sajjad@physics.harvard.edu}
\vspace{0.5cm}

{\it\small $^a$Jefferson Physical Laboratory, Harvard University,
Cambridge, Massachusetts 02138, USA}

\end{center}

\vspace{0.5cm}
\begin{center}
\today
\vspace{1cm}

\begin{abstract}

The s-confining theories are a class of supersymmetric gauge
theories with infrared dynamics which are well-understood. Perturbing
such theories can give rise to dynamical supersymmetry breaking. We
realize simple models of dynamical supersymmetry breaking by
perturbing two of the 10 $SU(N)$ s-confining gauge theories by a
single trilinear operator. These examples have locally stable vacua
with spontaneous supersymmetry breaking. The first is $SU(5)$ with two
generations (consisting of an antisymmetric tensor and an
antifundamental) plus two flavors. The second is $SU(5)$ with three
generations. The properties of the former vacuum are calculable while
those of the latter vacuum are not. We briefly discuss the other
$SU(N)$ models.


\end{abstract}

\end{center}

\maketitle

\newpage
\setcounter{page}{1}
\setcounter{footnote}{0}
\pagestyle{plain}

\section{Introduction and Overview}

Supersymmetry (SUSY) breaking at low scales is an attractive solution
to the hierarchy problem.  A compelling explanation for the origin of
such low scales is dynamical supersymmetry breaking (DSB), where small
mass scales arise from dimensional transmutation: $\Lambda =
M_{pl}\exp\left({{-8\pi^2 \over g^2(M_{pl})}}\right)$
\cite{Witten:1981nf}.  Understanding theories that exhibit this
structure requires computing non-perturbative or strong gauge
dynamics, which is in general very challenging.

Supersymmetric gauge theories with a weakly coupled magnetic dual
provide a tractable framework for building models of dynamical
supersymmetry breaking. In such theories, the effective description of
the infrared physics is severely constrained by global symmetries and
by the holomorphy of the superpotential. In particular, this technique
was used to understand vector-like $\mathcal{N} = 1$ theories
\cite{Affleck:1983rr,Affleck:1983mk,Seiberg:1993vc,Seiberg:1994bz,Seiberg:1994pq}
and later useful for understanding $\mathcal{N} = 1$ models with
vector-like matter and one antisymmetric tensor
\cite{Poppitz:1995fh,Poppitz:1996vh,Pouliot:1995me,Terning:1997jj,Craig:2011tx}.
Deforming these theories with $\delta W \not= 0$ can lead to models of
DSB, where the confining dynamics generates a weakly coupled
O'Raifeartaigh model \cite{O'Raifeartaigh:1975pr}. These models
typically possess an R-symmetry \cite{Nelson:1993nf} and relevant
operators with mass scales $m \ll M_{pl}$. For example, adding a tree-level mass perturbation, $\delta W =mTr(\bar{Q}Q)$, to vector-like $SU(N_c)$ supersymmetric QCD with $N_c
< N_F < \frac{3}{2}N_c$ flavors is sufficient to break supersymmetry
in a vacuum that is meta-stable but parametrically long lived
\cite{Intriligator:2006dd}. The O'Raifeartaigh model has an
approximate R-symmetry, and the hierarchically small mass scale is
given by $\sqrt{m\Lambda}$.

It is important to ask if other types of O'Raifeartaigh models can
emerge from a richer set of gauge dynamics. Happily, there exists a
special class of supersymmetric theories that exhibit confining
dynamics and are well understood, the so-called ``s-confining"
theories. In these theories, the moduli space has a smooth description
in terms of gauge invariant composites everywhere, including the
origin \cite{Csaki:1996sm,Csaki:1996zb}. The composites generally obey
classical constraint equations, which are realized in the IR theory by
the presence of a dynamical superpotential.  The classic example of an
s-confining theory is the case of $N_F = N_c + 1$ in the vector-like
$SU(N_c)$ gauge theories mentioned above. The gauge invariant
composites are the $N_F^2$ mesons, $M^{ij}= \bar{Q}^iQ^j$, the $N_F$
baryons, $B^i = \epsilon Q^{N_c}$, and the $N_F$ anti-baryons,
$\bar{B}^i = \epsilon \bar{Q}^{N_c}$ \cite{Seiberg:1994bz}. Below the
confinement scale $\Lambda$, the theory admits a description in terms
of the composites and a dynamical superpotential: \beq W^{dyn} =
\frac{1}{\Lambda^{2N_c -1}}\left( \bar{B}^iM^{ij}B^j - \text{det}{M}
\right).  \eeq As described above, a mass perturbation will lead to
supersymmetry breaking in this theory.

In this paper, we search for dynamical supersymmetry breaking vacua by
adding simple perturbations ($\delta W \not= 0$) to the $SU(N)$
s-confining theories.  See
\cite{Murayama:1995ng,terVeldhuis:1998jk,shadmi,singletassist} for
related investigations of some of the s-confining theories. A complete
list of the $SU(N)$ theories (as well as the $SO(N)$ and $Sp(N)$
theories) was found in \cite{Csaki:1996zb}, and is given in
Table~\ref{tab:SUN}. Our criteria for a simple perturbation is the
single lowest dimension operator allowed by gauge invariance, where we
will include weakly gauged global symmetries in our analysis in one
case.  We generally find three qualitatively different behaviors for
models obeying our criteria. First, most models possess no
supersymmetry breaking vacua, even locally. Second, some models have
calculable, locally stable, and supersymmetry-breaking vacua at small
field vacuum expectation values (VEVs). These vacua are parametrically
long lived and calculable, as in \cite{Intriligator:2006dd}. Third,
some models realize dynamical supersymmetry breaking, but in a vacuum
with field VEVs of order $\mathcal{O}(\Lambda)$; a controlled
calculation of the properties of the vacuum is not possible because
the \kahler\ potential is not known. However, one can conclude that
supersymmetry is broken based on the knowledge of the superpotential.

\begin{table}[bt]
\begin{center}
\begin{tabular}{|l|ll|} \hline
~~Theory~1 ~~& $SU(N)$ & $(N+1) (\Yfund + \overline{\Yfund})$  \\
~~Theory~2 ~~&$SU(2N)$ & $\Yasymm + 2N\, \overline{\Yfund} + 4\, \Yfund $  \\
~~Theory~3 ~~&$SU(2N+1)$ & $\Yasymm + (2N+1)\, \overline{\Yfund} + 4\, \Yfund $  \\
~~Theory~4 ~~&$SU(2N+1)$ & $\Yasymm + \overline{\Yasymm} + 3 (\Yfund + \overline{\Yfund})$ \\
~~Theory~5 ~~& $SU(2N)$ & $\Yasymm + \overline{\Yasymm} + 3 (\Yfund + \overline{\Yfund})$  \\
~~Theory~6 ~~&  $SU(6)$ & $\Ythreea + 4 (\Yfund + \overline{\Yfund})$  \\
~~Theory~7 ~~&$SU(5)$ & $ 3 (\Yasymm + \overline{\Yfund}) $  \\
~~Theory~8 ~~&  $SU(5)$ & $ 2\, \Yasymm + 2\, \Yfund + 4\, \overline{\Yfund}$ \\
~~Theory~9 ~~&$SU(6)$ & $2\, \Yasymm + 5\, \overline{\Yfund} + \Yfund$  \\
~~Theory~10 ~~&$SU(7)$ & $2 (\Yasymm + 3\, \overline{\Yfund})$  \\
  \hline 
\end{tabular}
\end{center}
\caption{s-confining $SU(N)$ theories \cite{Csaki:1996zb}. \label{tab:SUN}}
\end{table}

According to our criteria, most of the perturbed $SU(N)$ s-confining
theories do not break supersymmetry or are incalculable. We therefore
focus our analysis on two of the 10 models; our numbering of the
theories follows the convention of \cite{Csaki:1996zb}. Note that
theory 1 is precisely the s-confining example given above \cite{Intriligator:2006dd} and theory 6 is studied in \cite{shadmi}.  In Section~\ref{sec:theory8},
we study the effect of a trilinear perturbation to theory 8, an
$SU(5)$ gauge theory with two antisymmetric tensors, two fundamental
fields, and 4 antifundamental fields. A calculable and meta-stable
supersymmetry breaking vacuum exists near the origin of field space,
and the R symmetry is unbroken. In Section~\ref{sec:theory7} we
analyze theory 7, an $SU(5)$ gauge theory with three generations of an
antisymmetric tensor and an antifundamental field. Adding the simplest
trilinear perturbation gives rise to supersymmetry breaking, but with
a runaway direction. At large field VEVs, higher-dimension
perturbations can lift all flat directions, yielding a supersymmetry
breaking vacuum in an uncalculable regime. The approximate R-symmetry
is spontaneously broken. In Section~\ref{sec:theoryall}, we briefly discuss perturbations of
the other $SU(N)$ s-confining theories. 


\section{Theory 8 - $SU(5)$ with $2\, \protect \Yasymm + 4\ \overline{\protect\Yfund}+2\, \protect\Yfund$ \label{sec:theory8}} 

\begin{table}[bt]
\begin{displaymath}
\begin{array}{c|c|cccccc}
& SU(5) & SU(2)_1 & SU(4) & SU(2)_2 & U(1)_1 & U(1)_2 & U(1)_R \\ 
\hline
A & \Yasymm  & \Yfund & 1  & 1 & 0 & -1 & 0\\
\bar{Q} & \overline{\Yfund} & 1 & \Yfund & 1 & 1 & 1 & \frac{1}{3} \\ 
Q & {\Yfund} & 1 & 1 & \Yfund & -2 & 1 & \frac{1}{3} \\ 
\hline \hline
P_1 = Q\bar{Q} & & 1 & \Yfund & \Yfund & -1 & 2 & \frac{2}{3} \\
X = A\bar{Q}^2 & & \Yfund & \Yasymm & 1 & 2 & 1 & \frac{2}{3} \\
P_2 = A^2Q & & \Ysymm & 1 & \Yfund & -2 & -1 & \frac{1}{3} \\
Y = A^3\bar{Q} & & \Yfund  & \Yfund & 1 & 1 & -2 &\frac{1}{3} \\
Z = A^2Q^2\bar{Q} & & 1 & \Yfund & 1 & -3 & 1 & 1 \end{array}
\end{displaymath}
\label{tab:theory8}
\caption{The matter content and symmetries of Theory 8, an $SU(5)$
  gauge theory. $A, \bar Q$ and $Q$ are the microscopic degrees of
  freedom, and the other fields are the composites of the IR theory.}
\end{table}

Theory 8 has gauge group and matter content: $SU(5)$ with $2\,
\protect\Yasymm + 4\ \overline{\protect\Yfund}+2\, \protect\Yfund$,
where $\Yasymm$ is an antisymmetric tensor, $A$; $\overline{\Yfund}$
is an anti-fundamental, $\bar Q$; and $\Yfund$ is a fundamental,
$Q$. When $W = 0$, the theory has an anomaly-free $SU(2)_1 \times
SU(4) \times SU(2)_2 \times U(1)_1 \times U(1)_2 \times U(1)_R$ global
symmetry, with the charges given in Table~\ref{tab:theory8}.

In the strongly coupled IR regime, the theory has an s-confined
description in terms of five gauge invariant degrees of freedom: $P_1
= \bar{Q}{Q}, ~X = A\bar{Q}^2, ~ P_2 = A^2 Q, ~ Y = A^3\bar{Q}, ~ Z =
A^2Q^2\bar{Q}^2$. The global symmetry is unbroken, and there is a
dynamical superpotential which enforces the classical
constraints\cite{Csaki:1996zb,terVeldhuis:1998jk}:
\beq
W^{(dyn)} = \frac{1}{\Lambda^9} \left(  3XYZ - P_1^2Y^2 + 3P_1P_2XY- \frac{9}{32} P_2^2X^2  \right).
\eeq
We consider perturbing the $W = 0$ theory by the gauge invariant 
operator $A\bar{Q}^2$:
\beq
\label{t8pert}
\delta W = \lambda J_{iab}[A\bar{Q}^2]^{iab}
\eeq
where 
\beq
J_{iab} = {1 \over \sqrt{2}}\left[ \left( 
\begin{array}{cccc} 0 & 1 & 0 & 0 \\ -1 & 0 & 0 & 0 \\ 0 & 0 & 0 & 1 \\ 0 & 0 & -1 & 0 \end{array} \right), \left( 
\begin{array}{cccc} 0 & 1 & 0 & 0 \\ -1 & 0 & 0 & 0 \\ 0 & 0 & 0 & 1 \\ 0 & 0 & -1 & 0 \end{array} \right) \right],
\eeq
and $i \in SU(2)_1$, $a,b \in SU(4)$. This choice of perturbation
preserves an $Sp(4) \times SU(2)_2 \times U(1)\times U(1)_R$ symmetry,
where $Sp(4) \subset SU(4)$. Sitting inside these groups are four
anomaly free global $U(1)$'s. Gauging a linear combination of these is
sufficient to forbid all gauge-invariant operators up to and including
dimension 3, except for the non-zero terms appearing in
\eq{t8pert}. (This gauged $U(1)$ will also play an important role in
Section~\ref{sec:higgsing} when we consider the fate of the fields
$P_1$ and $P_2$.) Thus other allowed operators are higher dimensional
in the elementary description and suppressed by
$\mathcal{O}(\frac{1}{M^n_*})$, where $n$ is a positive integer and
$M_*$ is some high scale. We work in the limit that ${\Lambda \over
  M_*} \ll \lambda \ll 1$, in which case we can treat the operator in
\eq{t8pert} as a perturbation and neglect all other gauge-invariant
operators.

The s-confined description of the perturbed theory is then:
\beq
 W = \frac{1}{\Lambda^9} \left(  3XYZ - P_1^2Y^2 + 3P_1P_2XY- \frac{9}{32} P_2^2X^2  \right) + \lambda X.
\eeq
The \kahler\ potential is $K =
\frac{1}{\alpha_{P_1}|\Lambda|^2}P_1^{\dagger}P_1 +
\frac{1}{\alpha_X|\Lambda|^4}X^{\dagger}X +
\frac{1}{\alpha_{P_2}|\Lambda|^4}P_2^{\dagger}P_2 +
\frac{1}{\alpha_Y|\Lambda|^6}Y^{\dagger}Y +
\frac{1}{\alpha_Z|\Lambda|^8}Z^{\dagger}Z$, up to terms
$\mathcal{O}({\phi^{\dagger}\phi\phi^{\dagger}\phi})$ where
$\alpha_{\phi} > 0$ and $\phi = X, Y,Z, P_1,P_2$. These higher order
terms become negligible when studying the theory near the origin of
the moduli space, where $|\phi| \ll \Lambda$. Rescaling the fields for
canonical kinetic terms, we have
\beqa \label{sup8}
W &= & 3\sqrt{\alpha_X\alpha_Y\alpha_Z}XYZ + \lambda\sqrt{\alpha_X} \Lambda^2 X  \nn \\ &+&\frac{1}{\Lambda} \left( -\alpha_{P_1}\alpha_{Y}P_1^2Y^2 + 3\sqrt{\alpha_{P_1}\alpha_{P_2}\alpha_X\alpha_Y}P_1P_2XY- \frac{9}{32} \alpha_{P_2}\alpha_X P_2^2X^2\right) . 
\eeqa
The second line of the above equation consists of non-renormalizable
terms suppressed by $\Lambda$; these terms can be neglected if $|\phi|
\ll \Lambda$. We will return to this point later. Rewriting $h =
3\sqrt{\alpha_X\alpha_Y\alpha_Z}$ and $F = { \lambda \Lambda^2 \over
  3\sqrt{\alpha_Y\alpha_Z}}$, and putting in the explicit flavor
indices, the effective description of the theory near the origin
becomes:
\beq \label{t8mode}
W = h \left( \epsilon_{ij} \epsilon_{abcd} X^{iab} Y^{jc} Z^d +  F J_{iab} X^{iab} \right).
\eeq

We now show that near the origin, this theory admits a locally stable
supersymmetry breaking vacuum. We will show that supersymmetry is spontaneously broken, parameterize
the moduli space of the theory, choose a vacuum about which to expand,
and compute the 1-loop correction to pseudomoduli masses. We show that
the fields $P_{1,2}$ are also stabilized below $\Lambda$ due to the
presence of the weakly gauged flavor symmetry. This will suffice to
show that our choice of vacuum is locally stable. We then briefly
address the metastability of the vacuum.

First, this theory has no supersymmetric vacuum due to a rank-breaking
condition \cite{Intriligator:2006dd}. The F-term equations for the
field $X$ are:
\beq
  \left(F_X\right)_{1ab} = h F J_{1ab} + h \epsilon_{abcd} Y^{2c} Z^d ,\
  \left(F_X\right)_{2ab} = h F J_{2ab} - h \epsilon_{abcd} Y^{1c} Z^d.
\eeq
Here $J_{1ab}$ and $J_{2ab}$ are rank-4 antisymmetric
matrices. However, $\epsilon_{abcd} Y^{2c} Z^d$ is the antisymmetric
outer product of two vectors, $Y^{2c}$ and $Z^d$, which has a rank of
at most 2. Therefore there is no solution to the $F_X$ equations and
this is an example of rank-breaking. Furthermore, there is no runaway
direction with arbitrarily small potential energy.

We can find a vacuum in this theory by choosing field VEVs that cancel
off as many F-terms as possible, thereby minimizing the vacuum energy.
Since the rank of the antisymmetrized outer product of the $Y$ and $Z$
vectors is at most two, we can set two of the F-term equations to
zero by giving VEVs to the $Y$ and $Z$ fields. We have found solutions
by replacing $J_{iab} \rightarrow J'_{iab}$ where
\beq
J'_{iab} = {1 \over \sqrt{2}}\left[ \left( 
\begin{array}{cccc} 0 & 1 & 0 & 0 \\ -1 & 0 & 0 & 0 \\ 0 & 0 & 0 & 0 \\ 0 & 0 & 0 & 0 \end{array} \right), \left( 
\begin{array}{cccc} 0 & 1 & 0 & 0 \\ -1 & 0 & 0 & 0 \\ 0 & 0 & 0 & 0 \\ 0 & 0 & 0 & 0 \end{array} \right) \right].
\label{eq:theory8reduced}
\eeq
The solutions to these modified F-term equations give the
minimum-energy field configuration in the full supersymmetry breaking
theory.

Before we describe the fields VEVs in our vacuum, it will be
convenient to parameterize the fields $X,Y,Z$ in the following
way. Writing the field $X$ in terms of its $SU(2)_1$ components, $X^i
= (X^1, X^2)$, then we can define new fields
\begin{equation}
  X^A \equiv \frac{1}{\sqrt{2}} \left( X^2 - X^1 \right), \ 
  X^S \equiv \frac{1}{\sqrt{2}} \left( X^1 + X^2 \right), 
\end{equation}
where the fields $X^A$ and $X^S$ are
\begin{align}
 X^S \equiv & \left( \begin{array}{cccc}
          0 & x^S_{12} & x^S_{13} & x^S_{14}\\
	  -x^S_{12} & 0 & x^S_{23} & x^S_{24}\\
	  -x^S_{13} & -x^S_{23} & 0 & x^S_{34}\\
	  -x^S_{14} & -x^S_{24} & -x^S_{34} & 0
	  \end{array}
     \right), \ \
 X^A \equiv \left( \begin{array}{cccc}
          0 & x^A_{12} & x^A_{13} & x^A_{14}\\
	  -x^A_{12} & 0 & x^A_{23} & x^A_{24}\\
	  -x^A_{13} & -x^A_{23} & 0 & x^A_{34}\\
	  -x^A_{14} & -x^A_{24} & -x^A_{34} & 0
	  \end{array}
     \right), 
\end{align}
so that $X^i = \left( \left( X^S - X^A \right), 
          \left( X^S + X^A \right) \right)/\sqrt{2}.$
We make a similar field redefinition for the $Y$-fields: $Y^i = (Y^1,
Y^2)$, $ Y^A \equiv \left( Y^2 - Y^1 \right)/\sqrt{2},$ and $ Y^S
\equiv \left( Y^1 + Y^2 \right)/\sqrt{2}$, with
\begin{align}
 Y^S \equiv & \left( \begin{array}{cccc}
          y^S_1 & y^S_2 & y^S_3 & y^S_4
	  \end{array}
     \right), \ \
 Y^A \equiv \left( \begin{array}{cccc}
          y^A_1 & y^A_2 & y^A_3 & y^A_4
	  \end{array}
     \right), \ \ \nn \\
 & Y \equiv \frac{1}{\sqrt{2}} \left( \left( Y^S - Y^A \right), 
          \left( Y^S + Y^A \right) \right).
\end{align}
Finally, the $Z$ field is
\begin{displaymath}
 Z =  \left( \begin{array}{cccc}
          z_1 & z_2 & z_3 & z_4
	  \end{array} \right).
\end{displaymath}

The solution that minimizes the potential $V$ is
\begin{gather}
 z_1 = z_2 = y^A_1 = y^A_2 = y^S_1 = y^S_2 = x^S_{12} = x^S_{13} = x^S_{14} = x^S_{23} = x^S_{24} = x^A_{12} = 0, \nn \\
 \frac{x^A_{13}}{ x^A_{14}} = \frac{ x^A_{23}}{ x^A_{24}} = \frac{ y^S_3}{ y^S_4} =  \frac{z_3}{z_4} ~~~{\rm and}~~~~ y^A_3 = -\frac{1}{z_4} \left( F - y^A_4 z_3 \right).
\end{gather}
This set of equations can be generated by computing the
\groebner\ basis of the modified F-term equations from
\eq{eq:theory8reduced}.  The \groebner\ basis is a generating set for the ideal corresponding to a set of polynomial equations \cite{Buchberger:1976:TBR:1088216.1088219,MR1417938}, in this case the F-term
equations.  We used the \groebner\ basis algorithm implemented in
\texttt{Mathematica}.

We can use these equations to eliminate all variables but 8. These 8
complex variables span the classical moduli space:
\begin{equation}
  z_3, z_4, y^S_4, y^A_4, x^A_{14}, x^A_{24}, x^A_{34}, x^S_{34}.
\end{equation}
Therefore the classical vacuum structure of the renormalizable IR
theory, with the full perturbation to the superpotential, includes an
8-dimensional moduli space of vacua with the minimum potential energy
$V = 2 |h F|^2$.

At an arbitrary point in the moduli space, the fields are 
\begin{gather}
   X^A = 
     \left( \begin{array}{cccc}
          0 & 0 & x^A_{14} \frac{z_3}{z_4} & x^A_{14}\\
	  0 & 0 & x^A_{24}\frac{z_3}{z_4} & x^A_{24}\\
	  -x^A_{14} \frac{z_3}{z_4} & -x^A_{24}\frac{z_3}{z_4} & 0 & x^A_{34} \\
	  -x^A_{14} & -x^A_{24} & -x^A_{34}& 0
	  \end{array}
     \right), X^S = \left( \begin{array}{cccc}
          0 & 0 & 0 & 0 \\
	  0 & 0 & 0 & 0\\
	  0 & 0 & 0 & x^S_{34} \\
	  0 & 0 & -x^S_{34} & 0
	  \end{array}
     \right)
, \nn \\
 Y^A = \left( \begin{array}{cccc}
          0 & 0 & -\frac{F}{z_4} + y^A_4 \frac{z_3}{z_4} & y^A_4
	  \end{array}
     \right) , \ \ 
 Y^S = \left( \begin{array}{cccc}
          0 & 0 & y^S_4 \frac{z_3}{z_4} & y^S_4
	  \end{array}
     \right) , \ \ 
  Z = \left( \begin{array}{cccc}
          0 & 0 & z_3 & z_4
	  \end{array}
     \right)
\end{gather}

We pick a simple vacuum to expand about, with $z_4 = \sqrt{F}$ and all
other moduli set to zero:
\begin{align}
 X= 0, \ \
 Y = \sqrt{\frac{F}{2}}\left( \begin{array}{cccc}
          0 & 0 & 1 & 0\\
	  0 & 0 & -1 & 0
	  \end{array}
     \right) , \ \ 
  Z = \left( \begin{array}{cccc}
          0 & 0 & 0 & \sqrt{F}
	  \end{array}
     \right).
\end{align}

The light fields in this vacuum are either moduli or Goldstone bosons,
which we distinguish by examining the symmetries broken in this
vacuum. The $U(1)_R$ is unbroken because $X=0$. For the remaining $U(1)
\times Sp(4)$ symmetry, 4 linear combinations of the 11 generators are
preserved by the $Y$ and $Z$ VEVs, leaving 7 Goldstone bosons:
\begin{align}
  \text{Im} & \left( y^A_1 - \overline z_2 \right) , \  
  \text{Re}\left( \overline y^A_1 - z_2 \right) , \    
  \text{Im}\left( y^A_2 + \overline z_1 \right) , \  
  \text{Re}\left( \overline y^A_2 + z_1 \right) , \  \nn \\  
  & \text{Im}\left( y^A_4 + \overline z_3 \right) , \   
  \text{Re}\left( \overline y^A_4 + z_3 \right) , \  
  \text{Im}\left( y^A_3 + z_4 \right).
\end{align} 
As can be seen from the last 3 fields, three of the Goldstone bosons
mix with the moduli, in particular $y_4^A, z_3, z_4$. So the 16
classical moduli split into 13 pseudomoduli, which receive 1-loop
masses from supersymmetry breaking, and 3 Goldstone bosons, which
remain massless.

\subsection{One-loop masses}

We now evaluate the pseudomoduli masses at one-loop order and verify
that they are positive. The pseudomoduli $\phi$ are
\begin{equation}
  \phi \equiv ( z_3, \delta z_4, y^S_4, y^A_4, x^A_{14}, x^A_{24}, x^A_{34}, x^S_{34} )
\end{equation} 
with $z_4 = \sqrt{F} + \delta z_4$. As a check on our technique, we
have not parameterized away the Goldstone bosons and instead verify
that they remain massless when we compute the mass matrix for $\phi$.

The one-loop Coleman-Weinberg potential is \cite{Coleman:1977py}
\begin{align}
	V_{CW} =  \frac{1}{64 \pi^2} \str \left[ \emm^4 \log \frac{\emm^2}{\Lambda^2} \right]
\end{align} 
and the mass matrices are
\begin{equation}
	\emm^2_B = \left( \begin{array}{cc}
	\emm_F^* \emm_F & \eff^* \\
	\eff & \emm_F \emm_F^*
	\end{array} \right), \ \ \	\emm^2_F = \left( \begin{array}{cc}
	\emm_F^* \emm_F & 0 \\
	0 & \emm_F \emm_F^*
	\end{array} \right)
\end{equation}
where $\eff_{ij} \equiv F_k^* W_{ijk}$ and $\emm_F = W_{ij}$. Because
of the relatively large number of fields in the theory, it is not
straightforward to compute the eigenvalues of $\mb$ and $\mf$ as
functions of $\phi$.

It is therefore useful to have a simple numerical technique to
calculate the masses. For both fermion and scalar mass matrices, we
expand $\emm^2$ to quadratic order in the $\phi$ fields:
\begin{align}
	\emm^2 =  \text{max}\left(\emm^2(\phi_0) \right) \big[ \rm{I}   + \delta \emm^2(\phi) \large] + {\cal{ O}} (\phi^3),
\end{align}
where $\phi_0$ is the vacuum and $\text{max}\left(\emm^2(\phi_0)
\right)$ is the maximum eigenvalue of $\emm^2(\phi_0)$. Thus
$\text{max}\left(\emm^2(\phi_0) \right)$ is simply a number and the
matrix $\delta \emm^2(\phi)$ is a quadratic function of the fields
$\phi$. Dropping the ${\cal{ O}} (\phi^3)$ terms in $\emm^2$ (because
they will not be important for the masses), $V_{CW}$ can be expanded
as
\begin{align}
	V_{CW} =  \frac{1}{64 \pi^2} \str \left[ \emm^4 \log \frac{\text{max}\left(\emm^2(\phi_0) \right) }{\Lambda^2} + \sum_n \frac{(-1)^n}{n} \emm^4  \left( \delta \emm^2(\phi)  \right)^n \right] 
\label{eq:VCW_series}
\end{align} 
which is a straightforward calculation of products and traces of
matrices. The sum can be truncated at some order $n$, given a desired
numerical accuracy.

\begin{figure}
  \begin{center}
    \includegraphics[width=.6\textwidth]{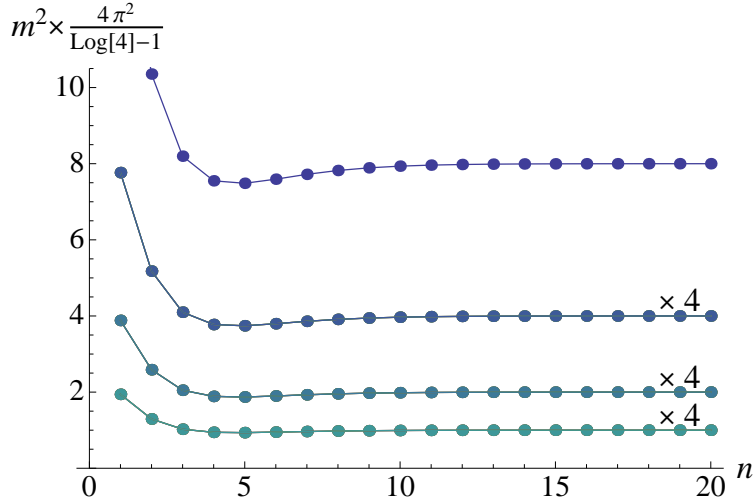}
    \caption{Pseudomoduli masses of Theory 8, in units of $|h^4 F|$, calculated to $n$th order in Eq.~\ref{eq:VCW_series}. The multiplicity of the lower three eigenvalues of indicated.}
    \label{fig:theory8masses}
  \end{center}
\end{figure}

The pseudomoduli mass matrix is:
\begin{equation}
  m^2_\phi= 
  \left( \begin{array}{cc}
    \frac{\partial^2 V_{CW}}{\partial \phi \partial \overline \phi} & 
       \frac{\partial^2 V_{CW}}{\partial \phi \partial \phi} \\
    \frac{\partial^2 V_{CW}}{\partial \overline \phi \partial \overline \phi} &
       \frac{\partial^2 V_{CW}}{\partial \phi \partial \overline \phi}
    \end{array}
    \right)_{\phi = \phi_0}
\end{equation}
We have calculated the mass eigenvalues for the pseudomoduli up to
$n=20$, and verified the convergence of the series expansion, as shown
in Fig.~\ref{fig:theory8masses}. There are always 3 massless
directions:
\begin{equation}
  \text{Im} (\overline y^A_4 + z_3),\ \text{Re}(y^A_4 + \overline  z_3),\ \text{Im}(\delta z_4)
\end{equation}
which are Goldstone bosons. The other 13 fields are true pseudomoduli
and have positive mass-squared values, which converge to the numerical
values of
\begin{equation}
   m^2_i = |h^4F| \left( 8, 4,4,4,4,2,2,2,2,1,1,1,1 \right) \times \frac{\log 4 -1}{4 \pi^2}.
\end{equation}
The structure of the problem is quite rigid; the eigenvectors are the
same at every order in the series expansion, as well as the ratios of
the eigenvalues. This seems to suggest there might be an analytic way
of resumming the series, as the potential may decouple into several
sectors as in \cite{Intriligator:2006dd}, but we do not pursue this
idea further.

\subsection{Higgsing pseudomoduli \label{sec:higgsing}}

The light fields $P_{1,2}$ appear only in $\Lambda$-suppressed terms
in $W$. The leading contribution to their masses, arising from both
the \kahler\ potential, $K$, and the superpotential, are of order
$\mathcal{O}(|h F/\Lambda|^2)$ and are thus incalculable without some
other stabilization mechanism.

However, if $P_1$ and $P_2$ have gauge interactions, they can feel
supersymmetry breaking more directly if the messengers are also
charged under the same gauge symmetry $G$. In this case, the relevant
term in the superpotential is
\begin{equation}
  W \supset 2 x^S_{34}  \left(F - z_1 y^A_2 + z_2 y^A_1 \right),
\end{equation}
so the messengers are $z_{1,2}$ and $y^A_{1,2}$. If the superpartner
to the goldstino, $x^S_{34}$, is neutral under $G$, then $P_{1,2}$
fall into the class of ``Higgsing pseudomoduli'' discussed in
\cite{Intriligator:2008fe}. The potential for $P_1$ and $P_2$ is
lifted at the two-loop level, at field values $|\sqrt{hF}| \ll
|P_{1,2}| \ll \Lambda$. While the masses of these pseudomoduli are
unknown at the origin, the fields are stabilized somewhere below
$\Lambda$.

The simplest candidate for $G$ is an anomaly-free subgroup of the
unbroken flavor symmetry of this model. For our purposes it is
sufficient to gauge the $U(1)$ associated with the generator
\begin{equation}
   a \sigma_z^2 + b \sigma_z^U  + c \sigma_z^D.
\end{equation}
Here $\sigma_z^2$ is a generator of $SU(2)_2$ and
$(\sigma_z^U,\sigma_z^D) \subset SU(2)^U \times SU(2)^D \subset
Sp(4)$.  This $U(1)$ is anomaly-free for any choice of the
coefficients $a, b,c$ above.
The charges of $P_1$ and $P_2$ are $q(P_2) = \pm a$ and $q(P_1) = \pm
a \pm b, \pm a \pm c$. The charges of the messengers are $q(z_{1,2}) =
\pm b$ and $q(y^A_{1,2} ) = \pm b$, and $x^S_{34}$ is
neutral. Therefore the $U(1)$ above satisfies the conditions for
``Higgsing pseudomoduli'' as long as $a$ and $b$ are nonzero, such
that $P_2$ and the messengers are charged. Note that gauging the
$U(1)$ above also justifies the previous statement that the
perturbation $ J A \bar Q^2 \sim x_{34}^{S}+ x_{12}^{S}$ is the lowest
dimension operator consistent with gauge-invariance that we could have
added to $W$.  The four components $x_{34}^{A,S}, x_{12}^{A,S}$ of the
$X$ field are neutral; the $SU(2)_1$ symmetry allows us to restrict to
only the fields $x_{34}^{S}, x_{12}^{S}$.

We now review the argument of \cite{Intriligator:2008fe} that Higgsing
pseudomoduli receive a 2-loop lifting of the superpotential away from
the origin. Suppose $|P_{1,2}| \gg |\sqrt{hF}|$, such that the effects
of supersymmetry breaking are small. Then the $U(1)$ gauge boson gets
a mass
\begin{equation}
  \frac{m_A^2(|P_1|^2,|P_2|^2)}{g^2} = \sum_i q( P_{1}^i)^2 | P_{1}^i|^2 + \sum_i q( P_{2}^i)^2 | P_{2}^i|^2,
\end{equation}
summing over all components of $P_1$ and $P_2$, and $g$ is the gauge
boson coupling. The gauge boson contributes to the anomalous
dimensions of the messengers, which include the term $-2 b^2 \frac{g^2
}{16 \pi^2}\subset \gamma$ above the scale $m_A$.

Because $x^S_{34}$ couples to the messengers, this introduces a
discontinuity of the $x^S_{34}$ anomalous dimension at the two-loop
level, which then generates the following effective potential in the
leading log approximation:
\begin{equation}
  V_{eff} (P_{1,2}) \approx 8 b^2 |h F|^2 \frac{h^2 g^2}{(16 \pi^2)^2}  \ln^2 \left( \frac{m_A^2(|P_1|^2,|P_2|^2)}{|h F| } \right).
\end{equation}
This lifts the $P_{1,2}$ directions\footnote{The quadratic part of the
  superpotential is invariant under a charge conjugation symmetry
  under which $z_1 \leftrightarrow y^A_2$, $z_2 \leftrightarrow
  y^A_1$, $x_{13}^S \leftrightarrow x_{24}^S$, and $x_{23}^S
  \leftrightarrow x^S_{14}$. This indicates that there is no one-loop
  D-term contribution to the $P_{1,2}$ masses. There will be
  contributions at higher loop, but these terms will be subdominant in
  the leading log approximation.}.

\subsection{Metastability}

The vacuum above was the lowest energy vacuum near the origin. In
general, the $\Lambda$-suppressed terms in $W$ could introduce a
supersymmetric vacuum somewhere in the moduli space. In fact, it is
possible to show that supersymmetry is still broken if these terms are
included; one way to see this is that the R-symmetry remains
unbroken. (See \cite{terVeldhuis:1998jk} for a similar analysis of the
full superpotential plus tree-level operators.) We have confirmed
there are no solutions to the the F-term equations, for the single
gauge invariant perturbation considered here, by again computing the
\groebner\ basis.  There could be higher dimension terms in $W$ as
well, but these would be suppressed by more powers of $\Muv$ and
generically don't restore supersymmetry at scales below $\Lambda$.

Other local minima may exist at field values at around $ \Lambda$,
though their vacuum energy is not calculable. If vacua with lower
energy exist, then the vacuum at the origin is metastable and can
decay. The decay rate can be computed by approximating the potential
as a square barrier; then one can use the results of
\cite{Duncan:1992ai}. Here the decay rate is $\propto \exp(-B)$, with
\begin{equation}
  B \sim \frac{ (\Delta \phi)^4 }{V_0} \sim \frac{\Lambda^4}{|h F|^2}. 
\end{equation}
The decay rate is exponentially suppressed for $|\lambda| \sim
|\frac{\Lambda^2}{hF}| \ll 1$, which is precisely the limit in which
our analysis is valid.


\section{Theory 7 - $SU(5)$ with $3(\protect\Yasymm +\overline{\protect\Yfund})$ \label{sec:theory7}}


\begin{table}[bt]
\begin{displaymath}
\begin{array}{c|c|cccc}
& SU(5) & SU(3)_1 & SU(3)_2 & U(1) & U(1)_R \\ \hline
A & \Yasymm  &  \Yfund  & 1 & 1 & 0 \\
\bar{Q} & \overline{\Yfund} & 1 & \Yfund & -3 & \frac{2}{3} \\
\hline \hline
X=A\bar{Q}^2 &1 & \Yfund & \overline{\Yfund} & -5 & \frac{4}{3} \\
Y=A^3\bar{Q} &1 & \Yadjoint & \Yfund & 0 & \frac{2}{3} \\
Z=A^5 & 1& \Ysymm & 1 & 5 & 0
\end{array}
\end{displaymath}
\label{tab:theory7}
\caption{The matter content and symmetries of theory 7, an $SU(5)$
  gauge theory. $A$ and $\bar Q$ are the microscopic degrees of
  freedom, and the other fields are the composites of the IR theory.}
\end{table}

$SU(5)$ with one generation (consisting of an antisymmetric tensor and
an antifundamental) was analyzed in
\cite{Affleck:1983vc,Murayama:1995ng}, and $SU(5)$ with two
generations was first analyzed in \cite{Affleck:1984uz}. Both cases
exhibited supersymmetry breaking when perturbations to $W$ were
added. To our knowledge, there has been no analysis of supersymmetry
breaking in an $SU(5)$ gauge theory with 3 generations, i.e., theory
7. With the aid of s-confinement, we now argue that this theory can
have non-supersymmetric vacua after including tree-level perturbations
to the superpotential.

Theory 7 has gauge group and matter content: $SU(5) ~{\rm with} ~3
\left( \Yasymm + \overline{\Yfund}\right)$, where $\Yasymm$ is an
antisymmetric tensor, $A$, and $\overline{\Yfund}$ is an
antifundamental, $\bar{Q}$.  When $W = 0$ the theory has an $SU(3)_1
\times SU(3)_2 \times U(1) \times U(1)_R $ anomaly free global
symmetry. The strongly coupled theory has an s-confined description in
terms of the three gauge invariants:
$X=A\bar{Q}^2,~Y=A^3\bar{Q},~Z=A^5$, with a dynamical superpotential
\cite{Csaki:1996zb}
\beq
 W^{(dyn)} = XYZ - \frac{20}{9}Y^3.
\eeq
The factor $\frac{20}{9}$ is determined by the classical constraint. 

We consider the effect of adding the three lowest dimension gauge
invariant operators to the superpotential:
\beq
\delta W = \lambda^{(1) i}_{I}[A\bar{Q}^2]^{I}_i+~\lambda^{(2)}_{ai} {\left[A^3\bar{Q}\right]^{ai} \over M_*} +  \lambda^{(3)}_{IJ}{\left[A^5\right]^{IJ} \over M^2_*},
\eeq
where $I~(a)$ is a fundamental (adjoint) index in $SU(3)_1$ and $i$ is
a fundamental index in $SU(3)_2$.  For generic values of
$\lambda^{(1),(2),(3)}$ one can use the techniques of
\cite{Luty:1995sd} to show that these perturbations lift all D-flat
directions. 

We will take $\frac{\Lambda}{M_*} = \eta \ll \lambda^{(1)}\sim
\lambda^{(2)} \sim \lambda^{(3)} \ll 1$, which is a natural hierarchy
when $\Lambda$ is generated by dimensional transmutation.  In this
limit, these interactions can be treated as perturbations of the $W=0$
theory. We first consider the effect of only the $\lambda^{(1)}
A\bar{Q}^2$ operator because it is the most important operator near
the origin of field space. Afterwards, we will address the effect of
$\lambda^{(2),(3)} \not= 0$.

An $SU(3)_1$ and $SU(3)_2$ flavor rotation can bring
$\lambda^{(1)i}_{I}$ into diagonal form. For simplicity, we assume in
what follows that $\lambda^{(1)}$ preserves the diagonal $SU(3)_D
\subset SU(3)_1 \times SU(3)_2 $,
\beq
\lambda^{(1)} =  \lambda \left( 
\begin{array}{ccc}  1 & 0 & 0 \\ 0 & 1 & 0  \\ 0 & 0 & 1 \end{array} \right).
\eeq
After rescaling the fields to obtain canonical kinetic terms, as
described in Section~\ref{sec:theory8}, the full ``s-confined''
superpotential becomes
\begin{equation} \label{sup7}
W =  h_2 \left( \frac{h_1}{h_2} X^{I}_i Y^{a i}T^{aJ}_{K}Z^{KL}\epsilon_{IJL} + \epsilon_{i j k}f^{abc}Y^{ai}Y^{bj}Y^{ck} + F \delta^i_{I}X^{I}_i \right)
\end{equation}
where $h_1 = \sqrt{\alpha_X\alpha_Y\alpha_Z}$ and $h_2 =
-\frac{20}{9}\alpha_Y^{\frac{3}{2}}$ are unknown dimensionless couplings; and $h_2
F = \sqrt{\alpha_X}\lambda \Lambda^2 \ll \Lambda^2$.
The Yukawa couplings $h_1$ and $h_2$ are marginally irrelevant. The
ratio of the couplings $r = \frac{h_1}{h_2}$ renormalizes to a fixed
point in the infrared: $\beta(r) = \frac{ (h_2)^2r}{32\pi^2}\left( -27
+ 7 r^2 \right)$ with two stable fixed points, $\beta(r_*) =0 $, at $
r_* = \pm 3\sqrt{\frac{3}{7}}$. Both interactions in the
superpotential are important for the IR physics.

The O'Raifeartaigh model of \eq{sup7} can be shown to have no
supersymmetric vacua at finite values for the field VEVs. We have
verified this by using the \groebner\ basis algorithm in
\texttt{Mathematica}, as described in Section~\ref{sec:theory8}.  If
the \groebner\ basis algorithm returns a constant as an element of the
generating set, then there is no solution at finite field
values. As in the previous section, we find the minimum reduction of
the full perturbation, such that there are solutions. In this case, the system of
equations has solutions if the rank of the matrix $\lambda^{(1)}$ is
reduced by one. An example of such a solution is
\beqa
~~~ \frac{Y^{ak}T^{aI}_{J}}{\sqrt{F}} &=& \left[ \left( \begin{array}
{ccc} 0&\frac{1}{\sqrt{2}} & 0 \\ \frac{1}{\sqrt{2}} & 0& -1\\ 0 & -1 & 0
\end{array} \right), \left( \begin{array}
{ccc} \frac{1}{\sqrt{2}} &0 & 1 \\ 0 & -\frac{1}{\sqrt{2}} & 0 \\ 1 & 0 & 0
\end{array} \right), 0 \right], \\ X^{I}_i &=& 0,   ~~~~~~~~~~~~~~ \frac{Z^{IJ}}{\sqrt{F}} = \frac{1}{r_*}\left( \begin{array}
{ccc} 0&0 & 0 \\ 0 & 0&0 \\ 0 & 0 & 1
\end{array} \right).
\eeqa

In the theory with a full rank $\lambda^{(1)}$ this vacuum breaks
supersymmetry, but is classically unstable due to the presence of a
runaway direction. This direction is parameterized by
\beqa
 \frac{Y^{ak}T^{aI}_{J}}{\sqrt{F}}&=& \left[ \left( \begin{array}
{ccc} 0&\frac{1}{\sqrt{2}} & 0 \\ \frac{1}{\sqrt{2}} & 0 & -1 \\ 0 & -1 & 0
\end{array} \right), \left( \begin{array}
{ccc} \frac{1}{\sqrt{2}} &0 & 1 \\ 0 & -\frac{1}{\sqrt{2}} & 0 \\ 1 & 0 & 0
\end{array} \right), \left( \begin{array}
{ccc} \frac{\epsilon}{6} & \frac{\epsilon}{2} & 0 \\ \frac{\epsilon}{2} & \frac{\epsilon}{6}  & 0 \\ 0 & 0 & -\frac{\epsilon}{3}
\end{array} \right) \right], ~~~\\ X^{I}_i&=& 0,~~~~~~~~~~~~~~~~~~~~~\frac{Z^{IJ}}{\sqrt{F}} = \frac{1}{r_*}\left( \begin{array}
{ccc} \frac{1}{\epsilon} &\frac{1}{\epsilon} & \frac{\sqrt{2}}{\epsilon} \\ \frac{1}{\epsilon} & -\frac{1}{\epsilon}& -\frac{\sqrt{2}}{\epsilon} \\ \frac{\sqrt{2}}{\epsilon} & -\frac{\sqrt{2}}{\epsilon} & 1
\end{array} \right).
\eeqa
The vacuum energy is $V \sim |\epsilon F|^2$, vanishing in the limit that $\epsilon \rightarrow 0$. Therefore the theory has a runaway direction in
the $Z$-fields.

Near the origin, the effects of the $\lambda^{(2)}$ and
$\lambda^{(3)}$ terms can be neglected. When $\langle Z \rangle \gg
\Lambda$, it is more convenient to describe the theory in terms of
elementary fields. In this case, the $\lambda^{(2)}$ and
$\lambda^{(3)}$ terms become important and the vacuum energy rises
with the VEV of the $Z$-field, as discussed earlier. We can then
conclude that a VEV for some field must develop around the scale $\Lambda$. A careful
analysis reveals that this theory possesses an approximate R-symmetry
where $R(X) = 2$, $R(Y)=-R(Z) = \frac{2}{3}$. Non-zero
values for $\lambda^{(2)}$ and $ \lambda^{(3)}$ introduce a small
explicit breaking of this R-symmetry. This model then falls into the classification of \cite{Goodman:2011jg}. On the runaway
direction, or for that matter anywhere except the origin, this
approximate R-symmetry becomes spontaneously broken since all fields
carry R charge. Therefore, we may conclude this theory spontaneously
breaks supersymmetry and also spontaneously breaks an approximate
R-symmetry at field values of $\mathcal{O}(\Lambda)$. However, the
details of this vacuum state are not calculable due to our ignorance
of the \kahler\ potential.


\section{Other $SU(N)$ s-confining theories} \label{sec:theoryall}

In this section we briefly describe perturbations to the other $SU(N)$
s-confining theories, focusing on the region of the moduli space near
the origin. After canonically normalizing the fields, we can neglect
$\Lambda$-suppressed interactions in $W$. This restricts the
interacting field content to fields that have relevant
interactions. We will only consider perturbing the $W=0$ theories with
a single lowest dimension operator consistent with gauge
invariance. The other $SU(N)$ theories generally fall into several
classes in this context.  Theories 2, 3 and 5 do not have
supersymmetry breaking for the simplest linear perturbation. There are
simple perturbations to the other theories which break
supersymmetry. However, theories 4, 6, and 10 possess irrelevantly
coupled pseudomoduli, which we find impossible to stabilize with a
gauged flavor symmetry. Meanwhile, theory 9 has a runaway in the
tree-level fields, similar to theory 7. As we discussed in detail in
Section~\ref{sec:theory7}, these theories generally have
supersymmetry-breaking vacua at field values of ${\cal{O}}(\Lambda)$,
but there is no small parameter with which to control the calculation
there.

In what follows, we omit ${\cal{O}}(1)$ factors and flavor indices for
simplicity and assume that kinetic terms have all been canonically
normalized.

\subsection{Theory 2 - $SU(2N)$ with $\protect\Yasymm +2N \ \overline{\protect\Yfund}
               + 4\ \protect\Yfund$}

Theory 2 is an $SU(2N)$ gauge theory with an antisymmetric tensor
$\protect\Yasymm\ (A)$, $2N$ antifundamentals
$\ \overline{\protect\Yfund}\ (\bar Q)$, and 4 fundamentals $
\protect\Yfund\ (Q)$.

\subsubsection{N$>$2}
For $N>2$, the IR theory has 3 singlets $X = A^N, Y = A^{N-2} Q^4,$ $Z
= \bar Q ^{2N}$ and an antisymmetric tensor under the $SU(4)$ flavor symmetry: $\Phi_i =
A^{N-1}Q^2$, where we have written $\Phi_i$ as a 6-vector for
convenience. The lowest dimension perturbation that can be added is $\delta W = h F X$. Because $X$ is a
singlet under the non-Abelian flavor symmetries, we cannot invoke a
gauge symmetry that forbids this operator but also permits other
perturbations. Thus the leading order superpotential is
\begin{equation}
  W = h( XYZ + Z \vec \Phi \cdot \vec \Phi + F X + ...)
\end{equation}
In this theory, supersymmetry is unbroken.

We note that supersymmetry is broken if we also include the next
lowest dimension perturbation, such that $\delta W = h( F X + \lambda
F \vec v \cdot \vec \Phi)$, with $\vec v = (0,0,0,0,0,1)$. However,
this theory possesses a runaway direction defined by, as
$\epsilon \to 0$,
\begin{equation}
  X = 0 , \ Y =  \sqrt{F} \epsilon,\ Z =  -\frac{\sqrt{F}}{\epsilon},\ \Phi = \lambda \sqrt{F} \epsilon \vec v.
  	\label{eq:theory2_runaway}
\end{equation}

\subsubsection{N=2}
In the $N=2$ case, the theory has a larger flavor symmetry, $SU(4)_1
\times SU(4)_2$, and an extra relevantly-coupled field $\chi_i =
A\bar{Q}^2$, an antisymmetric tensor under the $SU(4)_2$ flavor symmetry.
As in the $N >2$ case, the most general lowest-dimension operator we
can add is $\delta W = h F X$, and supersymmetry is unbroken.

We could consider higher order perturbations,
\begin{equation}
  W = h(XYZ +  Z \vec \Phi \cdot \vec \Phi+ Y \vec \chi \cdot \vec \chi  + F X + \lambda^{(1)}  F \vec v \cdot \vec \Phi  + \lambda^{(2)} F \vec v' \cdot \vec \chi + ...).
\end{equation}
Supersymmetry is still unbroken unless one of $\lambda^{(1,2)}$ is
zero; one can imagine doing this by weakly gauging a subgroup of one
of the $SU(4)$ flavor symmetries. Then supersymmetry is broken, but
there is a runaway direction. For the case $\lambda^{(2)} = 0$, the
runaway direction is given by Eq.~\ref{eq:theory2_runaway}, plus the
condition $\chi = 0$.

\subsection{Theory 3 - $SU(2N+1)$ with $\protect\Yasymm +(2N+1) \
               \overline{\protect\Yfund} + 4\ \protect\Yfund$}

The IR theory near the origin has an $SU(4)$ flavor symmetry, with an
$\Yfund \ (X^i),$ an $\overline{\protect\Yfund} \ (Y_i),$ and a
singlet $Z$. The lowest-order perturbed superpotential is
\begin{equation}
  W = h(\vec X \cdot \vec Y Z + F \vec v \cdot \vec X)
\end{equation}
where $\vec v = (0,0,0,1)$. Supersymmetry is unbroken in this case.

If the next order perturbation $\delta W = h \lambda F \vec v \cdot
\vec Y $ is also included, then supersymmetry is broken. However,
there is a runaway direction: $ Z \to \frac{\sqrt{F}}{\epsilon},\ \vec
Y \to - \sqrt{F} \epsilon \vec v,\ \vec X \to - \lambda \sqrt{F}
\epsilon \vec v$.

\subsection{Theory 4 - $SU(2N+1)$ with $\protect\Yasymm +\overline{\protect\Yasymm} + 3(\protect\Yfund +\ \overline{\protect\Yfund})$, $N=2$}

The full theory is $SU(2N+1)$ with $\protect\Yasymm
+\overline{\protect\Yasymm} + 3(\protect\Yfund
+\ \overline{\protect\Yfund})$. For simplicity we will only consider
the $N=2$ case. The lowest-dimension perturbation is $\delta W = h F
M_0$. Then then superpotential contains the terms
\begin{equation}
  W \supset h \left( (H_1)_a (\bar H_1)_{\bar b} M_0^{a \bar b} + F d_{a \bar b} M_0^{a \bar b}\right)
\end{equation}
where $a,\bar b$ are $SU(3)_1, SU(3)_2$ flavor indices and $M_0$
appears in no other terms in $W$. Thus this looks like an example of
rank-breaking. However, this theory contains a number of flavor
singlets $T$ which are only irrelevantly coupled, so the stability of
the vacuum near the origin is incalculable.

\subsection{Theory 5 - $SU(2N)$ with $\protect\Yasymm +\overline{\protect\Yasymm}+3(\protect\Yfund +\ \overline{\protect\Yfund})$,  $N=2$}

The theory is $SU(2N)$ with $\protect\Yasymm
+\overline{\protect\Yasymm} +3(\protect\Yfund
+\ \overline{\protect\Yfund})$, with flavor symmetry $SU(2) \times
SU(3)_1 \times SU(3)_2$.  Only the $N=2$ superpotential is given in
\cite{Csaki:1996zb}. The superpotential with the lowest dimension
perturbation is
\begin{equation}
   W = h(M_0 (M_2)^2 + H \bar H M_2 + F J M_0) 
\end{equation}
Here $H^i_a, \bar H^i_{\bar a}$ transform under $i \in SU(2), a \in
SU(3)_1, \bar a \in SU(3)_2$ while $(M_0)^{a\bar b}$ and $(M_2)^{a\bar
  b}$ transform under $SU(3)_1 \times SU(3)_2$. This theory has
supersymmetric vacua.

\subsection{Theory 6 - $SU(6)$ with $\protect\Ythreea +4(\protect\Yfund +
               \overline{\protect\Yfund})$}

This theory was recently investigated in \cite{shadmi}. The theory has
an $SU(4)_1 \times SU(4)_2$ flavor symmetry.  The lowest-dimension
perturbation is $\delta W = h F M_0$ and supersymmetry is broken via
rank breaking:
\begin{equation}
  W \supset h \left( (B_3)_a (\bar B_3)_{\bar b} M_0^{a \bar b} + F d_{a \bar b} M_0^{a \bar b}\right)
\end{equation}
where $a,\bar b$ are $SU(4)_1, SU(4)_2$ flavor indices and $M_0$
appears in no other terms in $W$. However, this theory contains a
singlet $T$ which is only irrelevantly coupled. 

\subsection{Theory 9 - $SU(6)$ with $ 2\, \protect\Yasymm +5\,  
               \overline{\protect\Yfund} +\protect\Yfund$}

The theory is $SU(6)$ with $ 2\, \protect\Yasymm \ (A) +5\,
\overline{\protect\Yfund} \ (\bar Q) +\protect\Yfund \ ( Q)$. The IR
fields are $M = Q\bar Q, X = A \bar Q^2, Y = A^3 Q \bar Q,$ and $Z =
A^4 \bar Q^2$ and the dynamical superpotential is
\begin{equation}
   W^{(dyn)} = M Z^2 + X Y Z
\end{equation}
with flavor symmetry $SU(2) \times SU(5)$. The lowest dimension
perturbation is $\delta W = h F J M$, but the theory does not break
supersymmetry.

We also considered the perturbation $\delta W = h F J X$, where
$X^{iab}$ is an $( \protect \Yfund, \protect \Yasymm)$ under $SU(2)
\times SU(5)$. The tensor $J$ is given by
\begin{equation}
  J_{iab} = \frac{1}{\sqrt{2}} \left( 
     \left( \begin{array}{ccccc}
          0 & 1 & 0 & 0 & 0\\
	  -1 & 0 & 0 & 0 & 0\\
	  0 & 0 & 0 & 1 & 0\\
	  0 & 0 & -1 & 0 & 0 \\
	  0 & 0 & 0 & 0 & 0 
	  \end{array}
     \right), \left( \begin{array}{ccccc}
          0 & 1 & 0 & 0 & 0\\
	  -1 & 0 & 0 & 0 & 0\\
	  0 & 0 & 0 & 1 & 0\\
	  0 & 0 & -1 & 0 & 0 \\
	  0 & 0 & 0 & 0 & 0 
	  \end{array}
     \right)
     \right).
\end{equation}
The perturbation is invariant under $Sp(4) \times U(1) \in SU(5)$;
weakly gauging this symmetry forbids the lower-dimension perturbation
$\delta W = h F J M$. This theory breaks supersymmetry, but not via
rank-breaking. We did not find a classically stable vacuum near the
origin and we expect there to be a runaway direction, analogous to
theory 7.

\subsection{Theory 10 - $SU(7)$  with $2\, \protect\Yasymm +6
\, \overline{\protect\Yfund}$}

This theory has already been discussed in \cite{Csaki:1996zb}. The
theory is $SU(7)$ with $2\, \protect\Yasymm\ (A) +6\,
\overline{\protect\Yfund} \ (\bar Q)$ and has IR degrees of freedom $H
= A \bar Q ^2, N = A^4 \bar Q$. The superpotential is $W^{(dyn)} = N^2
H^2 /\Lambda$, so there are no relevant interactions. It is possible
that this theory breaks supersymmetry if a perturbation in the $H$
fields is added. However, there is no calculable vacuum near the
origin.

\section{Conclusions}

Perturbations to $\mathcal{N} = 1$ supersymmetric gauge theories can
give rise to dynamical supersymmetry breaking. We have considered the
``s-confining" $SU(N)$ gauge theories of \cite{Csaki:1996zb} and added to the
superpotential a single gauge-invariant operator. These models are described by a
relatively complicated O'Raifeartaigh model near the origin of the
moduli space. We have employed two methods for analyzing the vacuum
structure. First, in order to solve the complicated F-term equations
that arise, it is useful to compute the \groebner\ basis. This also
gives a straightforward way of showing that supersymmetry is broken at
finite field values. Second, because of the large number of fields, we
have calculated 1-loop pseudomoduli masses numerically in a series
expansion. We have shown that the series converges for our example.

In our first model, we added a perturbation to theory 8, an $SU(5)$
gauge theory with two generations of an antisymmetric tensor plus
antifundamental and two flavors. Here we found a locally stable and
calculable vacuum with supersymmetry breaking near the origin of
moduli space. However, we were required to gauge an anomaly-free
subset of the global flavor symmetry. This ensured the stability of
the irrelevantly coupled pseudo-moduli, and justified our choice of
perturbation as the lowest dimension operator allowed by gauge
invariance.

The second example we gave was theory 7, an $SU(5)$ gauge theory with
3 generations of an antisymmetric tensor plus antifundamental. Our
analysis adds a new chapter to the history of $SU(5)$ models as
models of dynamical supersymmetry breaking. We found that adding the
lowest dimension gauge invariant operator removes any supersymmetric
vacuum from the theory. Near the origin, the framework of
s-confinement can be used to show that there is a runaway direction to
large field VEVs, at which point the elementary description becomes
valid. If there are other higher dimensional operators, which can be
neglected for the analysis near the origin, then all flat directions
can be lifted at large field values. The theory will stabilize in the
incalculable region with field VEVs of order
$\mathcal{O}(\Lambda)$. Furthermore, because we know that all fields
carry R-charge, the accidental R-symmetry is spontaneously broken.

We also briefly considered other $SU(N)$ s-confining theories and
found that, while they do have rich and interesting dynamics, they
either 1) do not lead to supersymmetry breaking when a single lowest
dimension operator is added to the theory, 2) they possess
irrelevantly coupled flavor singlets and cannot be rendered calculable
by gauging a flavor symmetry, or 3) they have tree level runaways to
large field VEVs, as in theory 7. In the latter two cases, any
supersymmetry breaking vacuum will generally be in an incalculable
regime.

\begin{acknowledgments}

We are grateful for many useful conversations with Frederik Denef,
Howard Georgi, Zohar Komargodski, Vijay Kumar, David Poland, Martin Schmaltz, Nathan Seiberg, Yael Shadmi, David
Shih, Yuri Shirman, and David
Simmons-Duffin. We thank Vijay Kumar and Yael Shadmi for
helpful comments on the paper. JM would like thank the Aspen Center
for Physics for their hospitality while this work was being done as
well as the organizers of the SUSY Breaking 2011 Workshop. This
research was supported by NSF Grant PHY-0855591.
\end{acknowledgments}

\bibliographystyle{h-physrev5}
\bibliography{refs}

\end{document}